# Self-Attention Assistant Classification of Non-Hermitian Phases in Two-Dimensional Lattice

Hengxuan Jiang, [1, §] Xiumei Wang, [2, §] and Xingping Zhou[3*]

[1] *College of Integrated Circuit Science and Engineering, Nanjing University of Posts and Telecommunications, Nanjing 210003, China*

[2] *College of Electronic and Optical Engineering, Nanjing University of Posts and Telecommunications, Nanjing 210003, China*

[3] *Institute of Quantum Information and Technology, Nanjing University of Posts and Telecommunications, Nanjing 210003, China*

*§ These authors contributed equally to this work.*

*\*zxp@njupt.edu.cn*

Classification of the non-Hermitian phases in high-dimensional lattice becomes challenging due to interplay of the band topology and non-Hermiticity. The significant increase in data dimensions and the number of categories has rendered traditional supervised learning and unsupervised manifold learning failed. Here, we propose the self-attention assistant machine learning for clustering non-Hermitian phases in two-dimensional lattice. By incorporating the self-attention mechanism, the model can effectively capture long-range dependencies and important patterns, resulting in a more compact and information-rich latent space. It can achieve Altland-Zirnbauer classification with Bloch vector dataset and distinguish the phases of eigenstates' localized behavior with the competition between non-Hermitian skin effect and topological localization. Our results provide a general method for characterizing non-Hermitian phases in two-dimensional lattice via machine learning.

*Introduction*. —The classification and identification of the transitions between different phases is a central task in non-Hermitian Altland-Zirnbauer (AZ) classification [1]. It is also a central task in the exploration of eigenstates localized behavior with the competition between non-Hermitian skin effect (NHSE) and topological localization. As non-Hermiticity arises naturally in a wide range of scenarios [2-7], it attracted tremendous attention in both theory [1, 8-24] and experiment [25-30]. The introduce of non-Hermiticity profoundly modifies the topological properties, leading to unprecedented phenomena beyond the descriptions of Bloch band theory, e.g., the breakdown of the conventional bulk-boundary correspondence [13, 14, 16-18, 23, 24]. Meanwhile, the competition between band topology and non-Hermicity makes the situation more elusive [31-42]. The introduction of the NHSE influences the distribution of topological phases significantly [31, 32]. Specifically, in non-Hermitian systems, the competition between band topology and non-Hermiticity in parameter space leads to two distinct scenarios for the system's eigenstates: symmetry-protected topological edge states and bulk-state localization driven by the non-Hermitian skin effect. Therefore, the localized behavior of eigenstates becomes an important indicator for distinguishing between non-Hermitian dominated regimes and topology dominated regimes. Since this competitive relationship affects the distribution of topological phases, it becomes essential to elucidate the interplay between band topology and non-Hermiticity.

Recently, machine learning has been adopted to solve challenges in classifying phases of matter and identify phase transitions. Within this vein, both supervised [43-53] and unsupervised learning [54-65] methods have been applied, enabling identifying different phases directly from complex data. However, despite the exciting progress made along this direction, its application suffers from some limitations as follows: (1) Supervised methods require the topological invariant as *prior* labels [43, 45, 48-52, 53]. It is still extremely challenging to find the topological invariant, with the introduction of non-Hermiticty, although the non-Bloch band theory [13, 14, 17, 18, 27, 66], biorthogonal bulk-boundary correspondence [8], and the non-Bloch Band theory in arbitrary dimensions [67] are proposed to solve this problem. Furthermore, the topological invariants in high dimensions exhibit intractable complexity to define. (2) A customized distance similarity function is required to handle specific topological models with periodic boundary conditions (PBCs) by unsupervised learning [57, 58, 61, 62, 65]. Some topological models, such as non-Abelian braiding, also require tailored similarity functions for effective clustering [68]. Thus, this method relies on the *prior* knowledge in physics. (3) It becomes tricky due to the possible existence of the NHSE with open boundary conditions (OBCs) [60, 69]. The competition between band topology and non-Hermiticity [31, 32, 36, 42] also leads to a more complex situation for the phases of eigenstates' localized direction. Those make the amount of random samples growing rapidly with the size of the Hilbert space. Due to the curse of dimensionality, it is difficult to deal with the model where various phases are present via machine learning.

In this work, we demonstrate an unsupervised learning method with a self-attention mechanism for the classification of the non-Hermitian phases in two-dimensional (2D) lattice. Inspired by the human cognitive system, computer scientists developed the self-attention mechanism [70-73]. It mimics the human ability to emphasize specific data features and ignore irrelevant parts. Therefore, we introduce this mechanism into the classification of the non-Hermitian phases in 2D lattice. Our method provides a global perception of the eigenstates, which does not rely on the *prior* knowledge, nor does it require customized distance similarity function. It can achieve non-Hermitian Altland-Zirnbauer (AZ) classification under PBC with Bloch vector dataset and distinguish the phases of eigenstates' localized behavior with the competition between NHSE and topological localization.

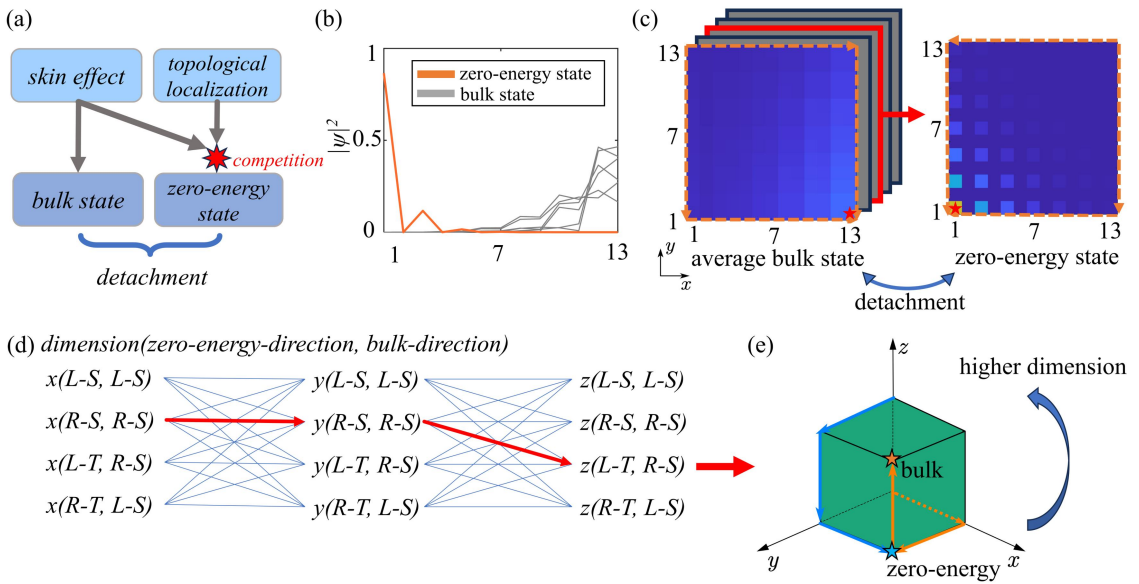

FIG. 1. The complex localized behaviors induced by the competition between NHSE and topological localization. (a) The detachment between the bulk and zero-energy states. (b) The 1D bulk state and zero-energy states in a non-Hermitian SSH model. (c) The 2D bulk state and zero-energy states in a non-Hermitian SSH model. The orange arrows represent the localized behaviors along the $x$ and $y$ directions. The red stars represent the localized positions of each eigenstate. (d) All the localized distribution in a 3D regular cube non-Hermitian SSH model with the detachment between zero-energy and bulk states. (e) The localized behavior of one 3D eigenstate.

*The detachment between bulk and zero-energy states.* —We first consider the detachment of zero-energy states and bulk states, which is a strong indicator of the competition between NHSE and topological localization [32, 33]. As is shown in Fig. 1(a), the localized behavior of the bulk states is dominated by the NHSE while that of zero-energy state is driven by the NHSE and the topological localization simultaneously. When topological localization overcomes the NHSE, the detachment between zero-energy states and bulk states occurs. At this time, the zero-energy states and the bulk states exhibit different localized distribution rather than aligned directions. It is shown in Fig. 1(c) that there is a detachment between the zero-energy and bulk states in the two-dimensional (2D) non-Hermitian SSH model. The NHSE can drive the bulk states to the right boundary. However, the zero-energy state is forced to the left boundary by topological localization, which overcomes the NHSE [36]. The detachment between the zero-energy and bulk states also occurs in high-dimensional non-Hermitian SSH models. There is a detachment between the zero-energy and bulk states in the two-dimensional (2D) non-Hermitian SSH model, shown in Fig. 1(c). As we can see, the 2D average bulk and zero-energy state are detached along the $x$ direction and aligned along the $y$ direction (see the orange arrow in Fig. 1(c)).

Next, we analyze the complex distribution of eigenstates induced by the detachment between the bulk and zero-energy states. For a high-dimensional Su–Schrieffer–Heeger (SSH) model, its eigenstates are originated from the one-dimensional (1D) eigenstates along each dimension [23, 35]. Thus, we list all the possible cases on each dimension to show the complex distribution of eigenstates on a higher dimension in Fig. 1(d). It shows the possible distribution of eigenstates in three-dimensional (3D) non-Hermitian SSH model. Each column refers the available distribution of the bulk and zero-energy state on the corresponding dimension. Each dimension supports four possible types of eigenstates, namely, left-localized skin states(L-S), right-localized skin states(R-S), left-localized topological states (L-T) and right-localized topological states (R-T). The notation $x$(L-S, R-S) denotes the L-S zero-energy state and R-S bulk states on $x$ dimension. The blue connection lines in Fig. 1(d) indicate the totally 4 distributions of localized behavior on 1D, 16 distributions on 2D and 64 distributions on three-dimensional (3D). In Fig. 1(e), we give the localized distribution of one 3D eigenstate, which corresponds to the combination represented by the red arrow. The yellow and blue stars represent the localized position of the 3D bulk and zero-energy state, respectively. The yellow and blue arrow shows the localized direction of the 1D eigenstate on the corresponding dimension. As the analysis shown in Fig. 1(d), each path across the three columns along the blue line corresponds to a distribution of eigenstates and homologous high-dimensional localized behaviors are diverse.

Thus, it is expected that more complex localized distributions will be induced by higher-dimensional competition between NHSE and topological localization. Furthermore, the classification algorithm needs to take all eigenstates into account due to the detachment between bulk and zero-energy states. As a result, the number of localized behaviors to be processed increases rapidly with the elevation of dimensions. Even more challengingly, the ~~number and~~ dimension of random samples grows rapidly with the Hilbert space as the size of the model increases.

The detachment between bulk and zero-energy states brings the conventional unsupervised learning algorithm a paradox: (1) Using a momentum-space dataset (PBC method) avoids the influence of finite size effects and the exponential large Hilbert space. However, the NHSE is overridden under PBC due to the sensitivity of boundary conditions [60]. Furthermore, the PBC methods heavily rely on the customized similarity distance function [68]. (2) Mapping the eigenstates in Hilbert space onto low-dimensional feature space (OBC method) provides a perceptual route to identify non-Hermitian phases and phase transition. However, the unsupervised learning of OBC method is particularly challenging due to the exponentially large high-dimensional Hilbert space. In addition, the OBC method relies on the customized similarity distance function as well [69].

To settle the problems above, we induce the self-attention assistant classifier to improve the representation for the extremely large number of high-dimensional samples. Our proposed self-attention assistant framework provides a global perception of eigenstates to adjust the attention for each of them. With the direct perception of eigenstates, the framework has the potential to solve all phase classification problems based on the localized behaviors of eigenstates.

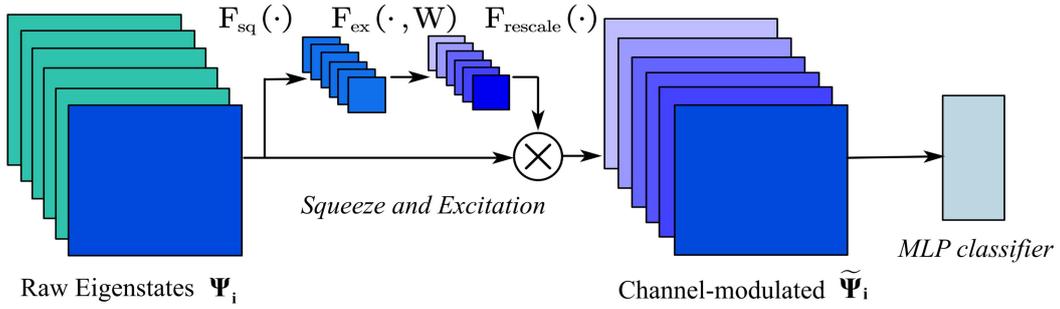

FIG. 2. The proposed self-attention assistant classification framework.

*The Dimensionality Reduction (DR).* —Most complex systems in nature conform to the low-rank hypothesis, meaning that despite their apparent high dimensionality, their intrinsic structure can often be captured in a much lower-dimensional space [79]. The DR is a non-linear transformation from raw eigenstates to their clustering-friendly latent representations. We apply Deep Clustering Network (DCN) for DR prior to clustering [74]. The DCN is a joint DR and K-means clustering framework that learns a much more clustering-friendly space by performing deep learning and clustering simultaneously. The DR part is implemented through the learning of a stacked autoencoder (SAE) and the clustering part is achieved by performing K-Means. The loss function of the DCN is:

$$\min_{\mathbf{W},\mathbf{Z},\mathbf{M},\{s_i\}} \sum_{i=1}^{N} (\frac{1}{N} | g(f(\mathbf{\Psi_i},\mathbf{W}),\mathbf{Z}) - \mathbf{X_i} |^2 + \frac{\lambda}{2} \| f(\mathbf{\Psi_i},\mathbf{W}) - \mathbf{M}s_i \|_2^2) \quad (1)$$

$$s.t. \quad s_{j,i} \in \{0,1\}, \quad 1^T s_i = 1 \quad \forall i,j \; .$$

We group the eigenstates at the same parameters as $\mathbf{\Psi_i} = \{|\psi_1\rangle, |\psi_2\rangle, \cdots |\psi_C\rangle\}$ and the eigenstates are sorted by the magnitude of their eigenvalues. The $C$ and $N$ represent the number of nodes in the model and the number of samples, respectively. The term $\frac{1}{N} | g(f(\mathbf{\Psi_i},\mathbf{W}),\mathbf{Z}) - \mathbf{X_i} |^2$ represents the reconstruction error of the SAE. The $f(\cdot,\mathbf{W})$ denotes the mapping function and $\mathbf{W}$ is a set of parameters that characterize the non-linear mapping. The $g(\cdot,\mathbf{Z})$ represents the decoding mapping of the SAE and $\mathbf{Z}$ is the parameters of the decoding network. The term $\frac{\lambda}{2} \| f(\mathbf{\Psi_i},\mathbf{W}) - \mathbf{M}s_i \|_2^2$ denotes the clustering error of the K-Means. The $s_i$ is the assignment vector for data point $i$ which contains only one non-zero element, $s_{j,i}$ denotes the $j$-th element of $s_i$, and the $k$-th column of $\mathbf{M}$, i.e., $\mathbf{m_k}$, denotes the centroid of the $k$th cluster. The $\lambda$ is a regularization parameter which balances the reconstruction error versus finding K-means-friendly latent representations [75]. The training process of DCN is divided into pre-training and formal training. The detailed workflow is provided in the Section I of the Supplemental Material [80]. In the pre-training stage, we only apply reconstruction loss to train the DCN (set $\lambda=0$). Then, the K-means is performed to obtain the initial $s_i$ and $\mathbf{M}$. The value $k$ is determined by the silhouette coefficient. The silhouette coefficient is given by [76]:

$$S = \frac{1}{N} \sum_{i=1}^{N} \frac{a_i - b_i}{\max(a_i, b_i)}$$
$$a_i = \frac{1}{n_A - 1} \sum_{j \neq i}^{n_A} d(l_i, l_j) \; . \quad (2)$$
$$b_i = \min_{L \neq A} d(l_i, L)$$

The $a_i$ and $b_i$ refer to the 'within' and 'between' similarity respectively. The $L = \{l_1, l_2, \ldots, l_N\}$ refers to the sample points in the latent space. The $A$ refers to the selected cluster and $n_A$ refers to the number of sample points in cluster $A$. The $d(\cdot,\cdot)$ refers to the Euclidean distance between two sample points.

The larger the silhouette coefficient is, the better the clustering results are. Based on this, we compute the silhouette coefficient *S(k)* with Eq. (2) for each value of *k* and select *k* when *S(k)* is as large as possible. Then, we start with the formal training by applying reconstruction loss and clustering loss simultaneously (set $\lambda \neq 0$). The latent representations are obtained after formal training. However, the eigenstates at the boundary of phase transition exhibit the localized behavior of two or more phases at the same time. It brings difficulties to the DCN to learn the segmentation between different phases. To optimize the segmentation, we induce the density filter.

*Density Filter.* —In the latent space, the distance between two sample points represents their similarity. The sample points with representative localized behavior gather into a density cluster, while the sample points at the phase transition boundary are sparsely distributed around the cluster. Based on this, we calculate the density gradient of sample points and only retain the sample points with the highest density. In this way, we are capable to select the points with the most representative feature.

We calculate the density of the sample points by Kernel Density Estimation (KDE). The density of the sample point *l* is given by:

$$D(l) = \frac{1}{n}\sum_{i=1}^{n} K(l - l_i). \tag{3}$$

The *n* refers to the number of sample points and the $K(\cdot)$ refers to the Gaussian kernel function. We set a threshold value of the density of points, below which the points are discarded. The remaining sample points are given the pseudo labels and are applied to train a classifier for topological classification.

*Self-attention Assistant Classifier.* —The mass localization features of eigenstates are simply repeated. Firstly, many bulk states share the same localized behavior. Secondly, the number of bulk states is much larger than the number of zero-energy state, which lead to the localization features of the zero-energy states being covered up. It implies that most of the features in the eigenstates are redundant for phase classification. To mitigate this problem, we induce an adaptive weight for each eigenstate to selectively emphasize informative features and suppress less useful ones

[73]. The adaptive weights are learned by capturing the long-range dependencies (the inter-eigenstate dependencies in there). The mechanism of such self-attention enables the classifier to capture the most salient localization features of the given eigenstates.

We apply the Squeeze and Excitation (SE) block [73] to our classifier. The SE block provides the global perception of the eigenstates and produces a collection of per-eigenstate modulation weights by learning the relationship between eigenstates. The mechanism of self-gating enables the SE block to capture the long-range dependencies and important patterns. As shown in Fig. 2, the SE block performs its functions through three basic operations: *squeeze*, *excitation* and *scale*.

The *squeeze* operation aggregates the eigenstates across their spatial dimensions and produces a descriptor $\mathbf{z} \in \mathbb{R}^C$ for all eigenstates. The $c$-th element of $\mathbf{z}$ is given by:

$$z_c = F_{sq}(|\psi_c\rangle) = \frac{1}{H \times W} \sum_{i=1}^{H} \sum_{j=1}^{W} \langle i, j | \psi_c \rangle \tag{4}$$

The given collection of eigenstates is denoted as $\Psi_i = \{|\psi_1\rangle, |\psi_2\rangle \cdots |\psi_C\rangle\}$, where $|\psi\rangle \in \mathbb{R}^{H \times W}$. Here, $H$ and $W$ represent the height and width of the eigenstate, respectively. $|x, y\rangle$ is the position basis of the wave function in 2D orthogonal coordinates. The function of this descriptor is to produce a low-dimensional embedding of the long-range (inter-eigenstate) information for the responses of the adaptive weights. The inter-eigenstate information is squeezed into low-dimensional space. The learning in this space gives the classifier the global perception of the eigenstates [73].

The *excitation* network is implemented by a bottleneck with two fully-connected (FC) layer. A ReLU function [77] and a sigmoid function are used to activate the first and second FC layer respectively. The network takes the eigenstate descriptor as the input, learns the nonlinear interactions between eigenstates and produces a set of per-eigenstate modulation weights to selectively gating the most salient feature. The weights $\mathbf{s} \in \mathbb{R}^C$ is denoted by:

$$\mathbf{s} = F_{ex}(\mathbf{z}, \mathbf{W}_{ex}) = \sigma(\mathbf{W}_2 \delta(\mathbf{W}_1 \mathbf{z})) \tag{5}$$

The $\delta$ and $\sigma$ refer to the ReLU function and sigmoid function respectively. The $\mathbf{W}_{ex}$, $\mathbf{W}_1$ and $\mathbf{W}_2$ refer to the weight of excitation network, the first FC layer and the second FC layer respectively. The *excitation* operation enables the network to fully capture the inter-eigenstate dependencies.

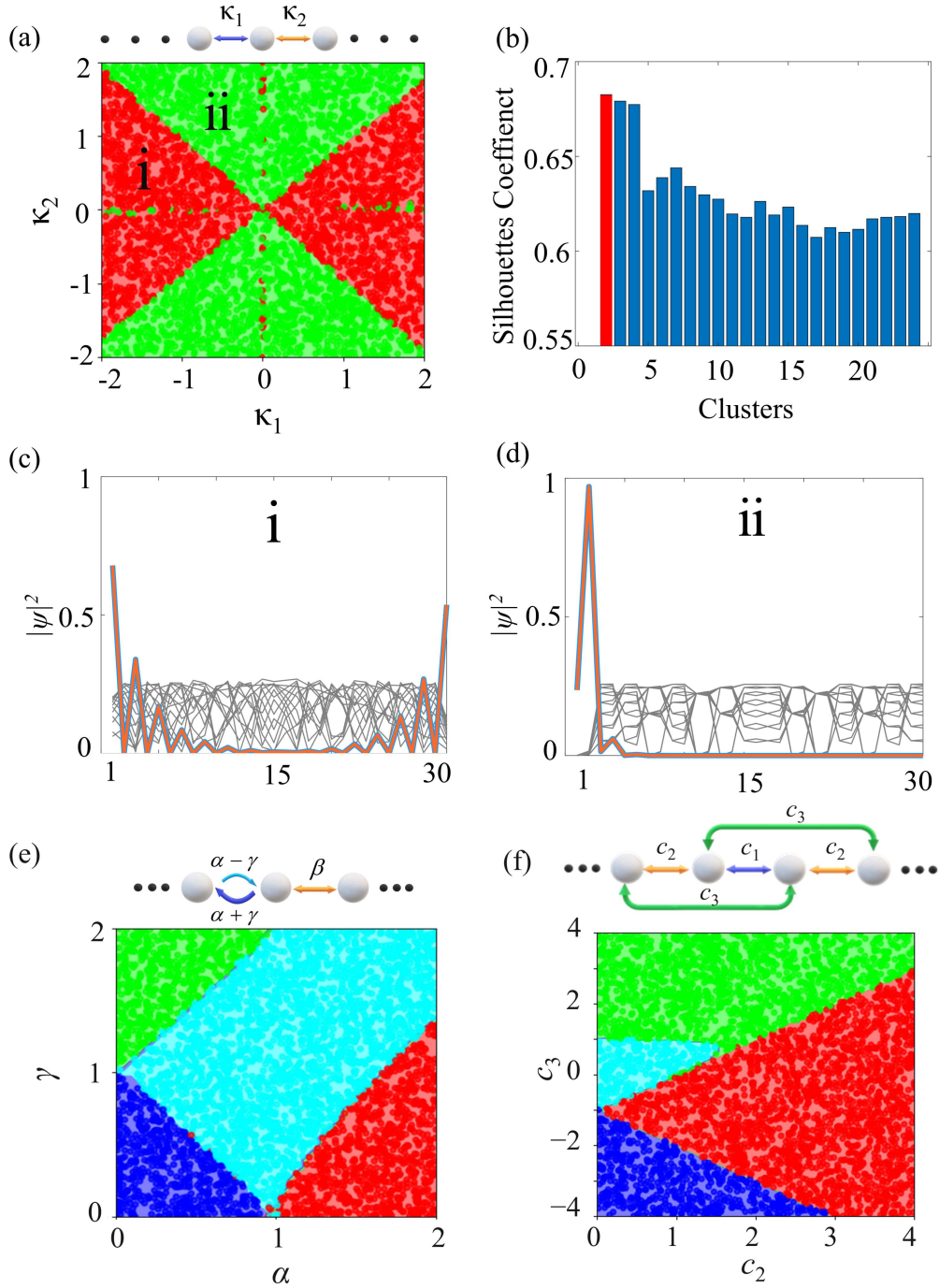

FIG. 3. The unsupervised classification of eigenstates' localized behavior and topological phases for the 1D SSH model. (a) The schematic illustration and learned phase diagram of the 1D SSH model for the eigenstates' localized behavior. (b) The numerical results of the silhouette coefficient after pre-training. (c) The specific eigenstates of classified phases i. (d) The specific eigenstates of classified phases ii. The value ranges of parameters $\kappa_1$ and $\kappa_2$ are both [-2, 2]. In the parameter space, 4000 points are randomly selected for the training of unsupervised learning.

(e) The phase diagram of AZ classification for 1D non-Hermitian SSH model (Eq. 7). (f) The phase diagram of AZ classification for 1D extended SSH model (Eq. 8).

The final output of the block is obtained by rescaling $\boldsymbol{\Psi}_i$ with the adaptive weight $s_c$:

$$|\widetilde{\psi}_c\rangle = \mathrm{F}_{scale}(|\psi_c\rangle, s_c) = s_c|\psi_c\rangle \tag{6}$$

where the modulated eigenstates $\widetilde{\boldsymbol{\Psi}}_i = \{|\widetilde{\psi}_1\rangle, |\widetilde{\psi}_2\rangle \cdots |\widetilde{\psi}_C\rangle\}$ and $\mathrm{F}_{scale}(|\psi_c\rangle, s_c)$ refer to multiplication between the scalar $s_c$ and the eigenstates $|\psi_c\rangle \in \mathbb{R}^{H \times W}$. We provide the detailed structure and parameters of proposed framework in the Section I and II of the Supplemental Material [80].

*The classification of the phases for eigenstates' localized behavior in 1D Hermitian SSH model.* — As shown in the Fig. 3(a), we first consider a 1D SSH model with the coupling coefficients $\kappa_1$ and $\kappa_2$. The sample points of input dataset are randomly chosen in the parameter space with varying $\kappa_1 \in [-2,2]$, $\kappa_2 \in [-2,2]$ and the corresponding eigenstates are concatenated into vectors with different channels. We feed all eigenstates with different parameters into our unsupervised method and perform the classification based on the eigenstates. The numerical results of the topological classification are provided in Fig. 3(a). The input samples are classified into 2 different topological -- phase i and ii, which are represented by the red and green region respectively. It is shown in Fig. 3(b) that the appropriate number of clusters is 2, as indicated by the numerical results of the silhouette coefficient. The eigenstates of phases i and ii are shown in the Fig. 3(c, d) respectively, where the orange lines represent the topological states. We find that the results of phase classification match exactly with the topological phase diagram obtained from the PBC method in 1D Hermitian SSH lattice.

*The AZ classification for 1D non-Hermitian SSH model.* - The AZ classification, originally developed for Hermitian systems, has been extended to non-Hermitian systems to account for their novel symmetries and topological structures. We have performed AZ classification under PBC for both a 1D non-Hermitian SSH model and an extended SSH model via our framework. The classification datasets for both models were constructed from their respective Bloch vectors (Eq. 7 for the non-Hermitian SSH model; Eq. 8 for the extended SSH model).

$$H_1(k) = \vec{d} \cdot \vec{\sigma} = (\alpha + \beta \cos k)\sigma_x + (\beta \sin k + i\gamma)\sigma_y,$$
$$\{\mathbf{x}_1^{(l_1)} \mid \mathbf{x}_1^{(l_1)} = [d(k_i), \mid k_i = [(2i - N - 2)/N]\pi, i \in (1, N)]\} \quad (7)$$

$$H_2(k) = \vec{d} \cdot \vec{\sigma} = (c_1 + c_2 \cos k + c_3 \cos 2k)\sigma_x + (c_2 \sin k + c_3 \sin 2k)\sigma_y,$$
$$\{\mathbf{x}_2^{(l_2)} \mid \mathbf{x}_2^{(l_2)} = [d(k_i), \mid k_i = [(2i - N - 2)/N]\pi, i \in (1, N)]\} \quad (8)$$

The randomly sampled points $l_1$ and $l_2$ are 4000 and 1600, respectively. The sampling points $N$ is set to 320 in the Brillouin zone. We employ a symmetric autoencoder architecture [320, 160, 80, 40, 20, 10, 2, 10, 20, 40, 80, 160, 320]. For the non-Hermitian SSH model, parameters $\alpha$ and $\gamma$ ranged within [0, 2] with $\beta$ fixed at 1. The results of classification show four distinct phases (Fig. 3(e)). For the extended SSH model, the parameters $c_2$ vary in [0, 4] and $c_3$ vary in [-4, 4]. The $c_1$ are fixed at 1. The result of classification exhibits four phases (Fig. 3(f)). These results validate the applicability of our framework. We also perform AZ classification for 1D non-Hermitian extended SSH model. The detailed results are provided in the Section VI of the Supplemental Material [80].

*The classification of the phase for eigenstates' localized behavior with 2D competition between NHSE and topological localization.* —We then consider the $13 \times 13$ 2D non-Hermitian SSH model as depicted in Fig. 4(a), of which the lower part shows the coupling coefficients $t_{1,x}$, $t_{1,y}$, $\delta$ and $t_2$. The 8000 sample points of input dataset are randomly chosen in the parameter space with varying $t_{1,x} \in [-2.4, 2.1]$, $t_{1,y} \in [-2.4, 2.1]$, while fixing $t_2 = 1.5$ and $\gamma = 0.2$. In this case, the dimension of the

samples **X** is given by $13 \times 13 \times 169 \times 8000 = 228,488,000$. The numerical results of the silhouette coefficient are depicted in Fig. 4(b), where the red column indicates sixteen as the appropriate number of clusters. As depicted in Fig. 4(c), the learned phase diagram shows the sixteen different phases ①-⑯. It is shown in Fig. 4(d) that the zero-energy states (left) and bulk states (right) exhibit reduced localized behavior at different phases. The table heads represent the 1D localized behavior of the eigenstates along the *x* and *y* directions respectively. As our analysis in Fig. 1(d), there are sixteen supported localized features for the 2D case. The specific localized behaviors of zero-energy and bulk states are shown in the Section III of the Supplemental Material [80]. In Fig. 4(d), we can see that the zero-energy states of phase ①, ④, ⑬ and ⑯ share the same localized behavior. However, the detachment of their bulk states forces them into different phases, which suggest the necessity of learning both zero-energy and bulk states simultaneously in proposed framework. What we would like to point out is our successful classification of phases ①, ②, ⑤, and ⑥. For these phases, the localized behavior of the zero-energy state serves as the most salient basis for classification, but they are obscured by a large number of bulk states with the same localization (the same applies to phases ③, ④, ⑦, ⑧, etc.). In conventional unsupervised learning, this problem becomes increasingly severe as the dimensions of the Hilbert space grow [69]. The proposed framework successfully overcomes this issue by selectively emphasizing the important patterns with self-attention mechanism. In Fig. 4(e), we show the specific localized behavior of the eigenstates at phase ③ and ⑨, of which the position is denoted by the red stars in Fig. 4(d). The left and right part of the Fig. 4(e) represent the localized behaviors of zero-energy and bulk states respectively. It can be seen that the classification of the phases matches exactly with our analysis in Fig. 1(d). To further show the validity of our method, we also apply the framework to the 2D non-Hermitian SSH model with a central defect. The detailed numerical results are provided in the Section IV and V of the Supplemental Material [80].

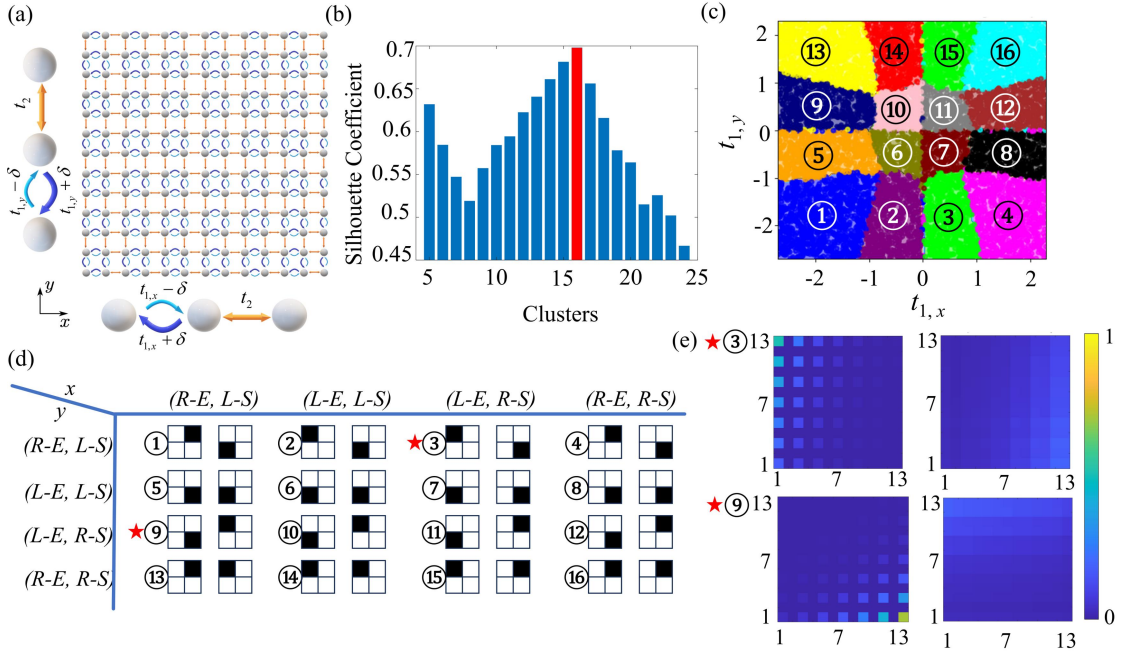

FIG. 4. The unsupervised classification of topological phase for the 2D non-Hermitian SSH model. (a) The schematic illustration of the 2D non-Hermitian SSH model. (b) The numerical results of silhouette coefficient after pre-training. (c) The learned phase diagram of 2D non-Hermitian SSH model. (d) The reduced localized behavior of the eigenstates in classified phases. The (L-E), (R-E), (L-S) and (R-S) represent left-localized edge state, right-localized edge state, left-localized skin state and right-localized skin state respectively. (e) The specific localized behavior of the eigenstates in phase ③ and ⑨. The value ranges of parameters $t_{1,x}$ and $t_{1,y}$ are both [-2, 2]. The $\delta$ and $t_2$ are fixed to 0.2 and 1.5, respectively.

*Conclusion.* —We have proposed an unsupervised learning of non-Hermitian phases in 2D lattice assisted by self-attention mechanism, which can effectively capture long-range dependencies and important patterns. The more compact and information-rich latent space obtained makes the classification process more efficient. Our method can be competent of handling classification tasks without a topological invariant nor a customized distance similarity function. Furthermore, we demonstrate the powerful capabilities in handling multiple topological phase classification. The proposed framework overcomes the issue where important eigenstate features are obscured due to the detachment between zero-energy and bulk states.-

In this work, our model addresses the AZ classification problem only under PBC.

The bulk spectra of non-Hermitian models can undergo dramatic changes according to the boundary conditions. As a result, the bulk Hamiltonian under OBC needs to be modified to restore the broken bulk-boundary correspondence, and the AZ classification results must also be adjusted accordingly [1, 19]. However, the transformation from a non-Hermitian Hamiltonian to a Hermitian Hamiltonian, which is the primary approach to obtaining the Hamiltonian under OBC, becomes particularly challenging in high-dimensional lattices. The amoeba theory [67] has already been established in a purely mathematical means for resolving the bulk-boundary correspondence problem in non-Hermitian systems of arbitrary dimensions.

In the context of unsupervised learning for high-dimensional space classification, we confront the persistent challenge known as the curse of dimensionality. This phenomenon manifests through an intriguing geometric paradox: as dimensionality increases, data points tend to concentrate near the surface of a hypersphere rather than being uniformly distributed, resulting in progressive loss of meaningful distance metrics and ultimately compromising classification efficacy [78]. To address this fundamental limitation, we have introduced the self-attention mechanism. This operational principle aligns with the mathematical foundations of the low-rank hypothesis [79], which posits that high-dimensional data inherently resides on lower-dimensional manifolds. Future advancements may require synergistic combinations of topological data analysis, and more subtle artificial intelligence algorithms to achieve robust high-dimensional pattern recognition.

**Acknowledgements**
The authors thank for the support by National Natural Science Foundation of China under (Grant 12404365). We thank Dr. Dandan Zhu for useful discussions.

# Supplementary Material for "Self-Attention Assistant Classification of non-Hermitian Phases in Two-Dimensional Lattice"

Hengxuan Jiang,[1, §] Xiumei Wang,[2, §] and Xingping Zhou[3*]

[1] *College of Integrated Circuit Science and Engineering, Nanjing University of Posts and Telecommunications, Nanjing 210003, China*

[2] *College of Electronic and Optical Engineering, Nanjing University of Posts and Telecommunications, Nanjing 210003, China*

[3] *Institute of Quantum Information and Technology, Nanjing University of Posts and Telecommunications, Nanjing 210003, China*

*§ These authors contributed equally to this work.*

*[*]zxp@njupt.edu.cn*

In this supplementary material, we will show additional supporting results for the detailed workflow of our proposed framework (Section I), the detailed parameters of our proposed framework used for the 2D cases (Section II), the specific localized behaviors of zero-energy and bulk states for 2D non-Hermitian SSH model (Section III), the classification of topological phase for 2D non-Hermitian SSH model with the central defect (Section IV), the specific localized behaviors of zero-energy and bulk states for 2D non-Hermitian model with central defect (Section V) and the classification result of non-Hermitian Altland-Zirnbauer classification (Section VI).

## I. THE DETAILED WORKFLOW OF OUR PROPOSED FRAMEWORK

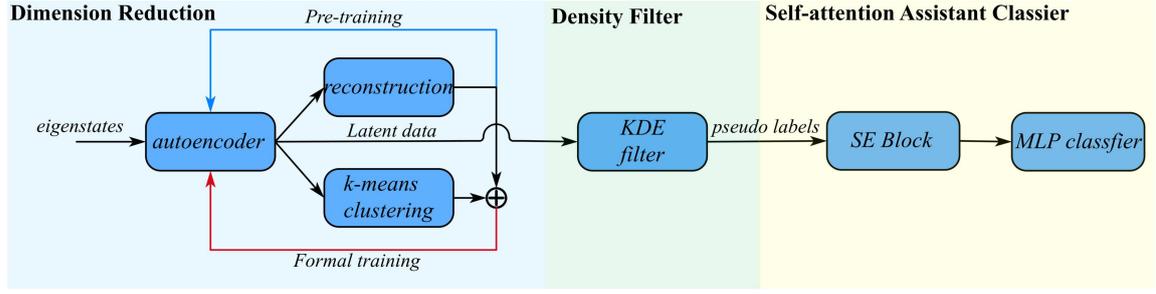

FIG. A1. The detailed workflow of our proposed framework.

As shown in Fig. A1, our framework is composed of three parts: the dimension reduction, the density filter and the self-attention Assistant Classifier. In the dimension reduction step, we first take the dataset of the raw eigenstates as input for the training of the Deep Clustering Network (DCN). In the pre-training stage, we only apply the reconstruction loss to the DCN by setting $\lambda=0$ in Eq. (1). In the formal training stage, we apply both the reconstruction loss and clustering loss by setting ($\lambda \neq 0$) in Eq. (1). After the training of DCN, we obtain the clustering-friendly latent representation of the dataset and the preliminary classification of the topological phases. Then, the sample points are filtered by density gradient in the latent space. We then attach the selected sample points with pseudo-labels and send them into the self-attention assistant classifier, which re-learn the features of the raw eigenstates using the selected sample points. Finally, we use the trained self-attention assistant classifier to generate the learned phase diagram.

## II. THE DETAILED PARAMETERS OF OUR PROPOSED FRAMEWORK USED FOR THE 2D CASES

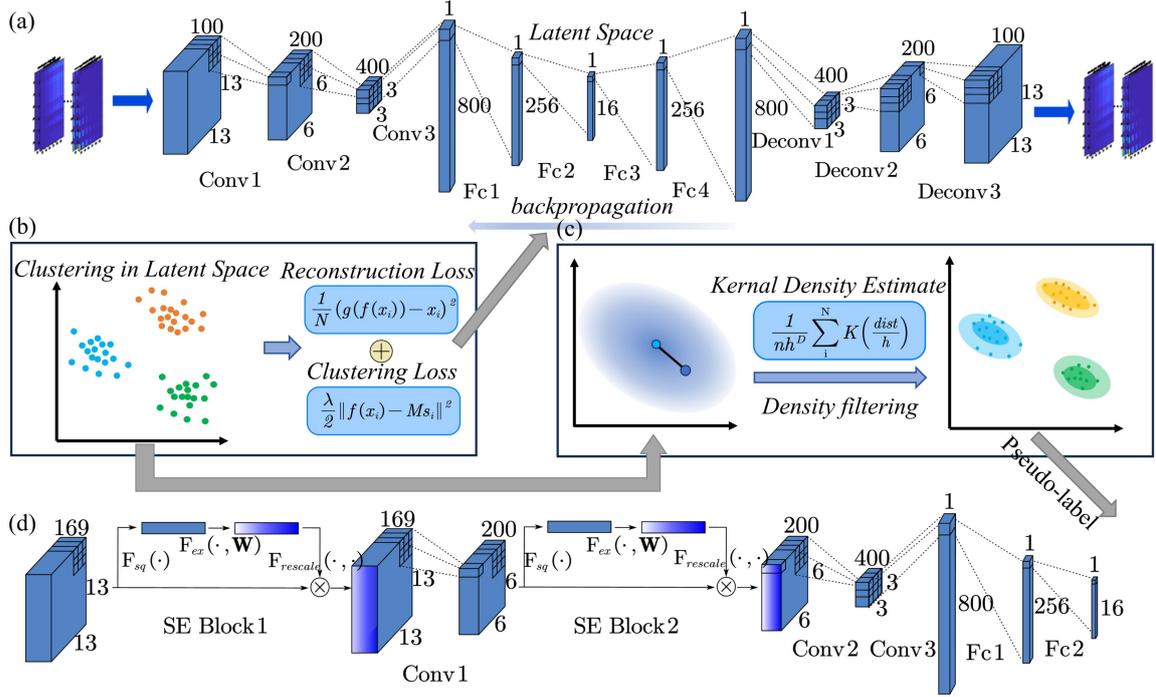

FIG. A2. The detailed parameters of our proposed framework used for the 2D cases. (a) The detailed structure and parameters of the convolutional autoencoder. (b) The illustration of the training process. (c) The illustration of the kernel density estimation and the density filter. (d) The detailed structure and parameters of the self-attention assistant classifier.

Fig. A2 shows the detailed parameters of the framework used for the topological classification of the 2D non-Hermitian SSH models. As shown in Fig. A2(a), we choose the convolutional autoencoder for the training of the DCN. The convolutional autoencoder consists of three convolutional layers, five fully connected layers and three deconvolution layers. The parameters of each layer are marked in Fig. A2(a). As depicted in Fig. A2(b), in the training process of the DCN, the self-supervised reconstruction of the autoencoder is carried out simultaneously with the k-means clustering and the loss function includes reconstruction error and clustering error. As shown in Fig. A2(c), we apply the kernel density estimation to calculate the density of sample points in the latent space and generate pseudo-labels. Fig. A2(d) shows the specific parameters of the self-attention assistant classifier. We apply one SE block in the frontend to extract the most salient features of the eigenstates for the classification and another SE block for improving the ability to capture the relations between eigenstates.

# III. THE SPECIFIC LOCALIZED BEHAVIORS OF ZERO-ENERGY AND BULK STATES FOR 2D NON-HERMITIAN SSH MODEL

We give the specific localized behaviors of zero-energy and bulk states for each learned phase in Fig. A3. Those match exactly with our classification results in Fig. 4(d). The parameters are as follows:

$t_{1,x} = -2.1$ and $t_{1,y} = -1.9$ for the phase ①, $t_{1,x} = -1.0$ and $t_{1,y} = -2.0$ for the phase ②, $t_{1,x} = 1.0$ and $t_{1,y} = -2.0$ for the phase ③, $t_{1,x} = 2.1$ and $t_{1,y} = -2.0$ for the phase ④, $t_{1,x} = -2.0$ and $t_{1,y} = -1.0$ for the phase ⑤, $t_{1,x} = -1.1$ and $t_{1,y} = -1.0$ for the phase ⑥, $t_{1,x} = 1.1$ and $t_{1,y} = -1.0$ for the phase ⑦, $t_{1,x} = 2.0$ and $t_{1,y} = -1.0$ for the phase ⑧, $t_{1,x} = -2.0$ and $t_{1,y} = 1.0$ for the phase ⑨, $t_{1,x} = -1.1$ and $t_{1,y} = 1.0$ for the phase ⑩, $t_{1,x} = 1.1$ and $t_{1,y} = 1.0$ for the phase ⑪, $t_{1,x} = 2.0$ and $t_{1,y} = 1.0$ for the phase ⑫, $t_{1,x} = -2.1$ and $t_{1,y} = 2.0$ for the phase ⑬, $t_{1,x} = -1.0$ and $t_{1,y} = 2.0$ for the phase ⑭, $t_{1,x} = 1.0$ and $t_{1,y} = 2.0$ for the phase ⑮, $t_{1,x} = 2.1$ and $t_{1,y} = 2.0$ for the phase ⑯. The $t_2 = 1.5$ and $\delta = 0.2$ remain the same for all the phases.

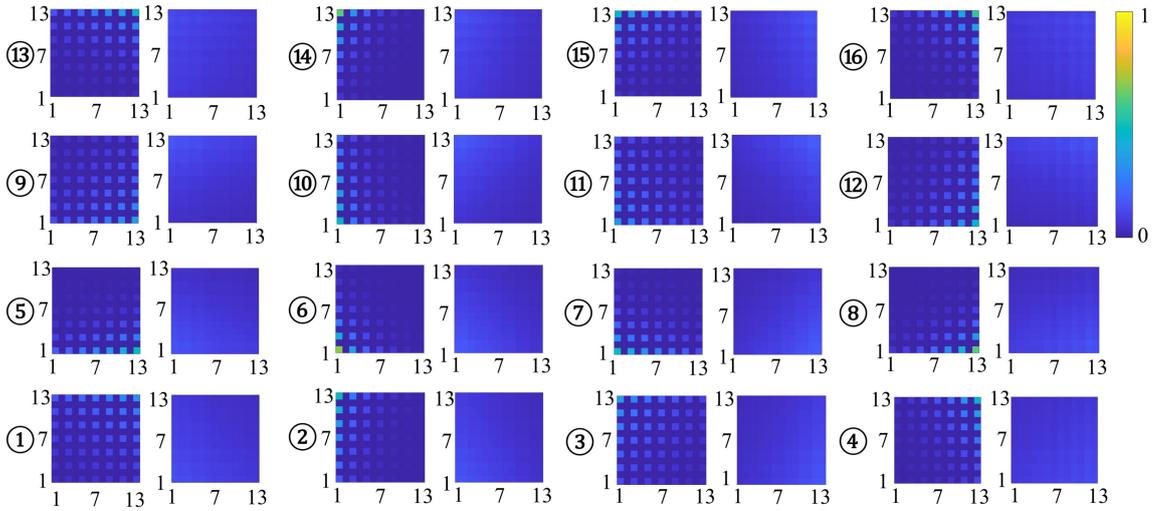

FIG. A3. The specific localized behaviors of zero-energy and bulk states for the 2D non-Hermitian SSH model.

# IV. THE SPECIFIC LOCALIZED BEHAVIORS OF ZERO-ENERGY AND BULK STATES FOR 2D NON-HERMITIAN SSH MODEL

We then consider the 2D non-Hermitian SSH model with the central defect as depicted in Fig. A4(a), of which the lower part shows the coupling coefficients $C_{1,x}$, $C_{1,y}$, $C_2$, $C_{n1}$ and $C_{n2}$. The sample points of input dataset are randomly chosen in the parameter space with varying $C_{1,x} \in [0,4]$, $C_{1,y} \in [0,4]$, while fixing $C_2 = 1$, $C_{n1} = 3.3$ and $C_{n2} = 0.5$. The numerical results of the silhouette coefficient are depicted in Fig. A4(b), where the red column indicates nine as the appropriate number of clusters. As depicted in Fig. A4(c), the learned phase diagram shows the nine different phases ①-⑨. Fig. A4(d) shows the reduced localized behavior of the zero-energy (left ones) and bulk states (right ones) at different phases. To simplify the notation, we denote the middle-localized topological state as (M-T). The table heads represent the 1D localized behavior of the eigenstates along the *x* and *y* directions respectively. In Fig. A4(e), we show the specific localized behavior of the eigenstates at phase ⑧ and ⑨, of which the position is denoted by the red stars in Fig. 4(d). The left and right part of the Fig. A4(e) represent the localized behaviors of zero-energy and bulk states respectively.

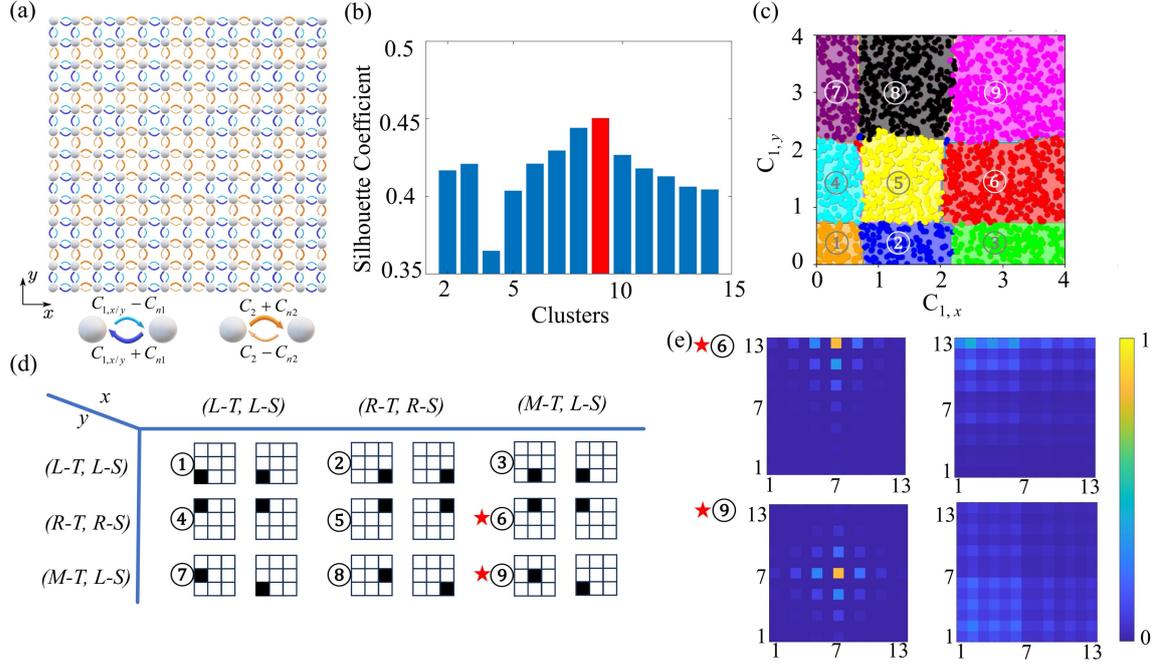

FIG. A4 The classification of topological phase for 2D non-Hermitian SSH model with the central defect. (a) The schematic illustration of the 2D non-Hermitian SSH model with the central defect. (b) The numerical results of silhouette coefficient after pre-train. (c) The learned phase diagram of 2D non-Hermitian SSH model with the central defect. (d) The reduced localized behavior of the eigenstates in classified phases. (e) The specific localized behavior of the eigenstates in phase ⑧ and ⑨.

# V. THE SPECIFIC LOCALIZED BEHAVIORS OF ZERO-ENERGY AND BULK STATES FOR 2D NON-HERMITIAN MODEL WITH CENTRAL DEFECT

We give the specific localized behaviors of zero-energy and bulk states for each learned phase in Fig. A5. Those match exactly with our classification results in Fig. A4(d). The parameters are as follows:

$C_{1,x} = 0.1$ and $C_{1,y} = 0.05$ for the phase ①, $C_{1,x} = 1.0$ and $C_{1,y} = 0.1$ for the phase ②, $C_{1,x} = 4.0$ and $C_{1,y} = 0.1$ for the phase ③, $C_{1,x} = 0.1$ and $C_{1,y} = 1.0$ for the phase ④, $C_{1,x} = 1.2$ and $C_{1,y} = 1.0$ for the phase ⑤, $C_{1,x} = 4.0$ and $C_{1,y} = 1.0$ for the phase ⑥, $C_{1,x} = 0.1$ and $C_{1,y} = 4.0$ for the phase ⑦, $C_{1,x} = 1.0$ and $C_{1,y} = 4.0$ for the phase ⑧,

$C_{1,x} = 3.9$ and $C_{1,y} = 4.0$ for the phase ⑨. The $C_2 = 1$, $C_{n1} = 3.3$ and $C_{n2} = 0.5$ remain the same for all the phases.

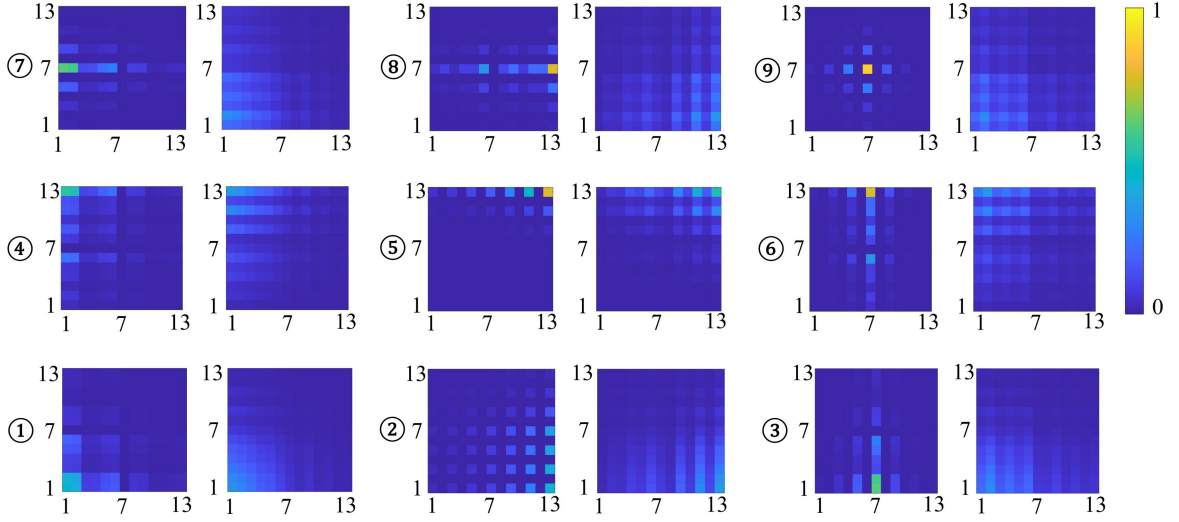

FIG. A5. The specific localized behaviors of zero-energy and bulk states in a 2D non-Hermitian SSH model with central defect.

## VI. THE RESULT OF NON-HERMITIAN ALTLAND-ZIRNBAUER CLASSIFICATION UNDER PBC WITH OUR UNSUPERVISED METHOD.

The Altland-Zirnbauer classification is originally developed for Hermitian systems and has been extended to non-Hermitian systems to account for new symmetry and topological structures introduced by non-Hermiticity. Our framework can be used for non-Hermitian AZ classification under PBC. We use the Bloch vector of the model as the input dataset. As shown in the upper part of Fig. A6, we select a one-dimensional non-Hermitian extended model with asymmetric coupling, and the dataset is represented as:

$$H(k) = \vec{d} \cdot \vec{\sigma} = (\alpha' + \beta' \cos k + \Delta \cos 2k) \sigma_x + (\beta' \sin k + \Delta \sin 2k - i\gamma') \sigma_y$$
$$\{\mathbf{x}^{(l)} \mid \mathbf{x}^{(l)} = [d(k_i)], k_i = [(2i - N - 2)/N]\pi, i \in (1, N)]\} \quad (A1)$$

where $l$ represents the number of randomly sampled points in the parameter space. $N$ denotes the number of sampling points in the Brillouin zone. $\alpha'$, $\beta'$, $\gamma'$ and $\Delta$ are the model parameters. In this classification task, the number of randomly sampled points $l$ is set to 4000 and $N$ is chosen as 320. The parameter $\alpha'$ is set within the range [0, 3]. The parameter $\gamma$ is set within the range [0, 3]. The parameter $\beta'$ and $\Delta$ are fixed at 1.

In the model framework, the autoencoder has the following architecture: [320, 160, 80, 40, 20, 10, 2, 10, 20, 40, 80, 160, 320]. The classification results are shown in the Fig. A6, where the input dataset is divided into six phases.

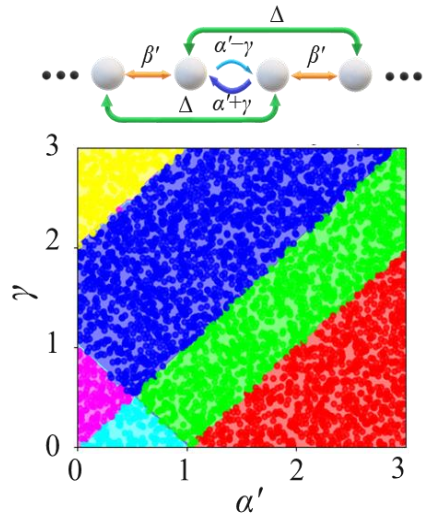

Fig. A6. The structure of 1D non-Hermitian SSH model and the results of non-Hermitian AZ classification under PBC with our method.